\documentclass[]{revtex4}
\usepackage{graphicx}

\def\bea{\begin{eqnarray}}
\def\eea{\end{eqnarray}}

\def\sqr#1#2{{\vcenter{\vbox{\hrule height.#2pt
      \hbox{\vrule width.#2pt height#1pt \kern#1pt
         \vrule width.#2pt}
      \hrule height.#2pt}}}}

\begin{document}
\title{ Has Hawking radiation been measured? }

\author{W. G. Unruh}
\affiliation{CIAR Cosmology and Gravity Program\\
Dept. of Physics\\
University of B. C.\\
Vancouver, Canada V6T 1Z1\\
~
email: unruh@physics.ubc.ca}
~

~

\begin{abstract}
It is argued that Hawking radiation has indeed been measured and shown to
posses a thermal spectrum, as predicted. This contention is based on three
separate legs. The first is that the essential physics of the Hawking process
for black holes can be modelled in other physical systems. The second is the
white hole horizons are the time inverse of black hole horizons, and thus the
physics of both is the same. The third is that the quantum emission, which is
the Hawking process, is completely determined by measurements of the classical
parameters of a linear physical system. The experiment conducted in 2010
fulfils all of these requirements, and is thus a true measurement of Hawking
radiation.
\end{abstract}

\maketitle
\section{Introduction}
\label{intro}
In 1974 Hawking \cite{hawking} predicted one of the most surprizing phenomena in
gravitational physics, and possibly in physics in general. That prediction was
that black holes, objects whose spacetime structure was such that no radiation
could propagate, even in principle, from inside the object to an outside
observer, nevertheless produced radiation which gradually shrank the size of
the black hole. Furthermore, that outgoing radiation had the spectrum of black
body radiation, modified by an "albedo" factor. The temperature of the
radiation for a non-rotating charged black hole was given by 
\bea
T= {\hbar c^3\over Gk_B}{1\over 8\pi M}
\eea
where $k_B$ is Boltzmann's constant, $G$ is Newton's gravitational constant,
$\hbar$ is Planck's constant, $c$ the velocity of light, and $M$ the mass of
the black hole.  

That the so called albedo was just that, and not a frequency dependent
temperature, was demonstrated by showing that a black hole connected to a heat
bath well outside the black hole would be in equilibrium if that outside
temperature were given by that temperature. (This equilibrium is the so called Hartle
Hawking ``vacuum"). 

That black holes could radiate was the first shock. Where did these emitted
particles come from? They could not come from inside the black hole, since
nothing can travel faster than light  and even light  cannot escape from
inside. If
they come from outside the horizon, exactly what creates them out there? The
second shock  was
that this radiation was thermal.  What causes this temperature? Does this mean
that black holes are thermodynamic objects, like other hot objects? Do they
have entropy, and what is the relation of the entropy of black holes to other
forms of entropy? Is the second law of thermodynamics valid when  this entropy is
taken into account? 

All of these questions have been some  of the foremost topics in theoretical
physics in the years since Hawking's result, and are  questions which still do not have 
universally accepted answers. 

However, if one examines Hawking's original calculation, there are some severe
problems with his derivation. While mathematically unimpeachable, they are nonsense physically.
The reason is intimately tied to the fact that nothing can escape from a black
hole. Therefor, if one looks at that emitted radiation, and asks where it must
have come from, since it is travelling away from the black hole now, it must have
been closer to the black hole in the past. But it cannot have been inside the
black hole. The equations for quantum field theory, used to predict the
radiation, say that as time unwinds into the past, that radiation must have
been closer and closer to the horizon, squeezed into a shorter and shorter
distance, and thus a shorter and shorter wavelength. This is an exponential
process, so that the wavelength decreases exponentially with the time into the
past, with a time scale crudely set by the light crossing time of the black
hole (i.e., the time taken for light to travel a distance equal to the
circumference of the black hole).  This process continues until finally one
arrives at the time in the past when the black hole formed, presumably by the
collapse of matter. That radiation then came, through the centre of the
collapsing star, from the space outside the collapsing star. 

Thus, if one follows Hawking's calculation and one looks for the origins of that
thermal radiation in the behaviour of the field in the distant past, the past
aspects of the quantum field which creates the current radiation had 
wavelengths of order ${GM\over c^2}e^{-t{c^3/4GM}}$ and frequencies of order
${c^3\over GM}e^{t{c^3/4GM}}$. Thus one second after a solar mass black hole forms, the
radiation, produces by whatever the process is that produces Hawking
radiation, originated from frequencies in the initial state of the uncollapsed
system of order $e^{10^5}$, a number so absurdly large that any imaginable
units would simply produce an insignificant change in that exponent. And the
later the radiation one is considering is emitted, the larger and more absurd
this factor becomes. 

There is simply no way that the physical assumptions-- namely that the
quantum field which produces this radiation propagates linearly on the
background unaltered spacetime-- can be correct. Those frequencies which are
needed to explain the radiation produced even one second after a solar mass
black hole forms, correspond to energies which are  $e^{10^5}$ times the
energy of the whole universe. Such waves simply will not propagate with no
effect on the background spacetime and will certainly not propagate as though
the black hole were unaffected by its presence. That these fluctuations are
"vacuum fluctuations" should make no difference to this observation.

The question thus arises-- if the derivation relies on such absurd physical
assumptions, can the result be trusted? If the physics of the emission process
really does depend on the physics of the field at those frequencies, then
surely one can regard the effect as at best highly speculative, and  and at
worst almost
certainly wrong. 

When it was discovered, this process was seen to be something unique to black
holes. Without a complete theory of quantum gravity, it would seem that one
could not make any progress toward understanding this process. But in 1980,
while teaching a course in fluid mechanics, I realised that there might be
another way of approaching the problem. Many waves, including sound waves in a
fluid, have a behaviour at low frequencies and long wavelengths which is almost
identical to that of relativistic fields in a spacetime. Already in the
1920's, Gordon\cite{gordon} had realised that at low frequencies and long wavelengths,
sound waves obey equations which obey  a "special relativity" set of 
transformations of space and time. If the background fluid were forced to
flow, then that background flow would alter the equations of motion of the
sound waves in precisely the same way that a non-flat spacetime metric would
alter the equations of motion for fields in the spacetimes corresponding to
Einstein's theory of gravity\cite{unruh-dumbhole}. In particular, if one modelled the equations of
motion of sound waves as an irrotational ($\nabla\times v_1=0$ where $v_1$ is
the first order perturbation of the flow away from the background flow $v$) perturbation of
the fluid, then the velocity potential defined by  $v_1=\nabla\phi$ obeyed
exactly the equations of motion of a scalar field in a metric
\bea
{1\over \sqrt{|g|}} \partial_\mu \sqrt{|g|} g^{\mu\nu} \partial_\nu \phi=0
\eea
where, in the case of the fluid, the metric coefficients were given by 
\bea
\sqrt{|g|} g^{\mu\nu} = \rho \left(\begin{array}{cc}-c^2& v^i\\ v^j
\delta^{ij}-v^i v^j\end{array}\right)
\eea
Here  $g$ is the inverse of the determinant of the matrix $g^{\mu\nu}$, and
$c$ is the velocity of sound in the fluid $\sqrt{{\partial p\over \partial\rho}}$ ( which may depend on position and
time). 

Since one can easily imagine the fluid  somewhere flowing faster than the velocity of
sound, sound waves from inside the surface on which the "radial"
velocity of the fluid is equal to the velocity of sound cannot escape that
region, just as light cannot escape the black hole. The metric, whose
components are the inverse of the matrix $g^{\mu\nu}$, can contain a horizon
which is the exact analog of the horizon of a black hole. 

\begin{figure}\begin{center}
    \includegraphics[width=80mm]{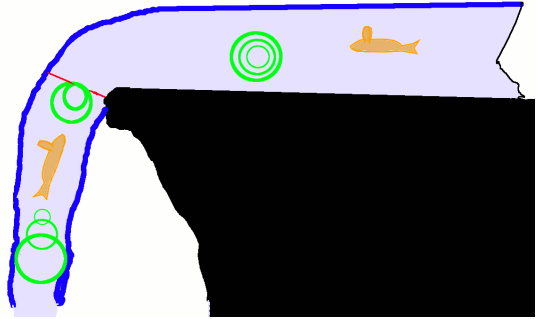}
    \caption{ The sound waves emitted by a yelling fish as it goes
over a waterfall which goes supersonic at the red line. Just as for a black
hole, beyond the sonic horizon, the sound waves are swept over the falls with
the fish. (Note that the appendages on the heads of the fish are ears, not
fins, since these fish experience the world through sound, not sight}
    \end{center}\end{figure}

By following Hawking's derivation, line by line, for a fluid flow which
accelerates to create such a ``horizon", one predicts that the quantum sound
waves in such a fluid flow should also create quantum particles around the
horizon, which should again have a temperature, in this case proportional to 
\bea
T= {\hbar\over k_B} {1\over 2\pi c} {\partial (c^2-v^2)\over \partial r} 
\eea
evaluated on the surface where $c^2= v^2$\cite{unruh-dumbhole}\cite{visser} .

Again, if one remains in the hydrodynamic approximation, the derivation
suffers from the same difficulties as does that for the black hole radiation,
namely that the radiation appears to depend on absurdly high frequencies and
short wavelengths in the initial state of the system.  

Unlike for gravity, however, for fluids we understand the short wavelength, high
frequency physics, at least in principle. Fluids are made of molecules, and
once the wavelength of the sound waves becomes comparable to the distance
between the molecules, the hydrodynamic approximation fails. The equation of
motion of the fluid particles are no longer continuum equations, but become
finite difference type equations (assuming we can neglect special relativistic
effects). While at wavelengths much longer than the inter-atomic spacing,
continuum, field theory type approaches are valid, at short wavelengths they
no longer suffice. It was recognized by Jacobson\cite{jacobson}  that one of the key effects
that this atomicity had was on the dispersion relation of the small
fluctuations about some equilibrium flow of the fluid. In a fluid at rest, the
relation between the frequency and wavelength was no longer the simple
\bea
\nu\lambda=c
\eea
where $\nu$ is the frequency and $\lambda$ the wavelength, but $\nu$ has a
much more complex relation to $\lambda$. 
\bea
\nu= F\left({1\over \lambda}\right)
\eea
where F is some potentially complicated function of ${1\over\lambda}$ such that
at large $\lambda$ $F$ becomes a linear funtion with slope $c$. The phase
velocity $F\lambda$ and group velocity
$-\lambda^2{\partial F\over\partial\lambda}$ both will differ from $c$ for
small $\lambda$. One can thus take a first step at understanding the
dependence of the thermal radiation on the nature of the theory of the waves
at short wavelengths by examining the behaviour of the prediction under
changes in the dispersion relation of the waves at short wavelengths. 

In a fluid with such a dispersion relation changes, we can again ask "What
aspect of the state of the fluid in the past results in the thermal nature of
the radiation emitted now?"  The outgoing wave-packet projected
back from the future is that as time unrolls into the past, the packet gets
closer to the horizon, and  its wavelength
decreases just as in the back hole model. However,eventually its wavelength becomes small enough that the dispersion
relation, and thus the group velocity of the wave changes. That wave-packet can
no longer stay near the horizon (where the velocity of the fluid is now
different from the changed velocity of the wave). As one goes back further
into the past, that wave packet must have come from
either inside the horizon (if the dispersion relation is such that the
group velocity of the waves increases as the wavelength decreases) or outside (if
the group velocity decreases as the wavelength decreases). This stops the
exponential change in wavelength that the horizon brings about. The system has
a natural cutoff to the decrease in the wavelength caused by the horizon. The
radiation emitted now comes from aspects of the quantum field (the sound
field) in the past which have much shorter, but not absurdly shorter than the
wavelengths now.  The wavelengths are not determined by an exponential of the
time since the formation of the horizon, but rather by the dispersion
relation. It is when the dispersion relation changes from the long wavelength,
relativistic, regime, to the cutoff regime where the atomicity of matter
becomes important. 

A variety of numerical studies (eg, early ones are Unruh\cite{unruh-numerical}
and Corley and Jacobson\cite{corley-jacobson} but see the references in
Barcelo {\it et al} Living Reviews article \cite{visser})  have shown that this system will still emit thermal
quantum radiation. Changes in the short wavelength dispersion relation have no
(or only a very small) effects on the temperature or thermal spectrum of the radiation
emitted. 
\begin{figure}\begin{center}
    \includegraphics[width=80mm]{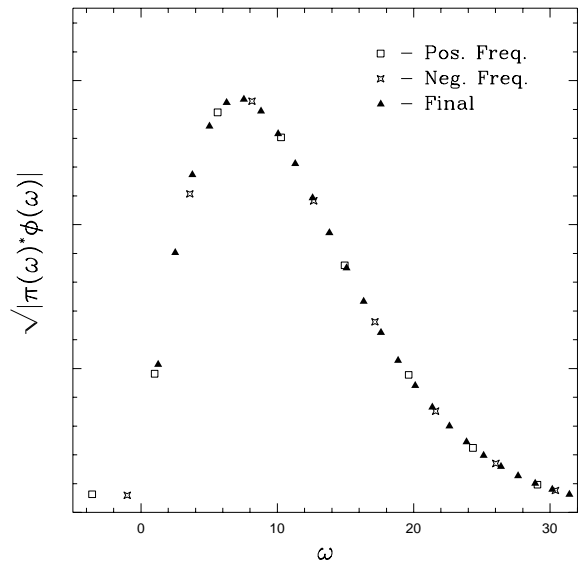}
    \caption{ The numerical result of of calculating the the thermal
factor for a dispersive horizon. If the waves obey the thermal hypothesis, the
three curves should be the same, which they are (except possibly at the
longest wavelengths where the size of the calculation region is or the order
of the wavelength, and the calculation is unreliable.) \cite{unruh-numerical}}
    \end{center}\end{figure}

However, as always in physics, experiments are the final arbiter. Do other
physical effects (viscosity in the fluid, turbulence, etc.) alter that thermal
spectrum?

Since $\hbar$ is so small, quantum effects, like Hawking's prediction, will
always be very small. Is it possible to measure such effects? For fluids, with
typical laboratory velocities of meters per second, and changes on the scale
of cm., the temperature of the radiation would be expected to be of the order
of  ${\hbar\over k_B}{\Delta v\over \delta x}\approx 10^{-10}K$. This is
clearly extremely difficult to measure directly. However, as Hawking's
calculation already showed, the quantum emission follows directly from the
classical behaviour of the system. Nowhere, except at the very end his
calculation, did quantum mechanics play any role. 

For any linear system, the classical and the quantum behaviour, and classical
equations 
and the quantum  Heisenberg equations of motion are identical. It is only the
interpretation of the symbols that occur in the calculation that differ.
Instead of the field values, $\phi$ and conjugate momentum $\pi$ being
interpreted as ordinary function, having some distinct real value at a point
in spacetime, those symbols represent linear operators operating on a Hilbert
space. Those operators obey non-trivial commutation relations, and it is those
commutation relations that differentiate the classical and the quantum
systems. 

\section{Experiment}
\label{exper}

In 2010, a group at the University of British Columbia (Silke Weinfurtner, Ted
Tedford, Matt Penrice, Greg Lawrence, and I-- a group of theoretical
physicists and civil engineers)\cite{experiment} carried out an
experiment to measure the spectrum of radiation produced by a horizon in an
analog system following a suggestion of Sch\"utzhold and
Unruh\cite{schuetz-gravitywave}. The system of interest was water flowing in a flume ( a long,
narrow tank down which water flows). We had a 6m long tank, with a width of
about 15cm, down which water, with a depth of about 20cm flowed. At the
outflow end the water fell into a large storage tank. From this tank the water was
pumped to the other end of the flume, where it entered the flume through a
pipe and flowed through a progression of screens. These screen were to smooth
out the flow of the water, so that the flow in the rest of the tank was as
laminar as possible. ( the screens convert gross turbulence into small scale
turbulence which is rapidly damped out by the viscosity of the water. 

Along the flume is a smooth, aeroplane wing shaped obstacle on the bottom which
forces the water to flow more rapidly over the top of the obstacle. The exit
slope of the obstacle was adjusted to ensure that there was no stagnation in
the flow as it left the obstacle. The flow was measured using  Particle Image
Velocimetry (in which neutrally buoyant particles are introduced into the
water and their velocity measured by taking high resolution photographs of the
particles illuminated by two closely spaced laser flashes) to ensure that
flow separation did not occur. This showed that although the flow was laminar
throughout, as the flow can down the trailing edge of the obstacle there was a
decrease of the velocity with depth to about 50\% of the velocity along the
top of the stream. 

\begin{figure}\begin{center}
    \includegraphics[width=80mm]{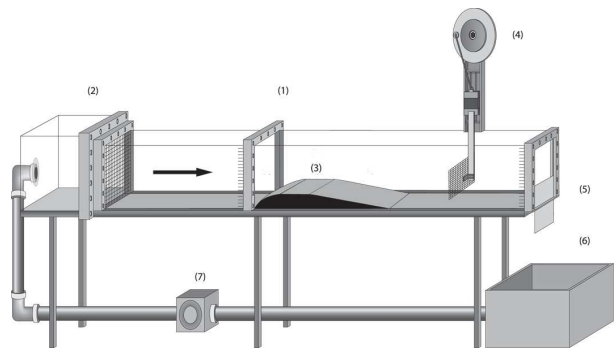}
    \caption{ Diagram of the flume with the obstacle and the wave
generator. The water falls over the weir at the end and is recirculated by the
pump. The laser light is a narrow sheet 2 meters long and about 1mm wide along
the centre of the flume from the top of the obstacle
downstream.\cite{experiment} }
    \end{center}\end{figure}

In the experiment, we were interested in measuring the surface waves along the
top of the flow. These surface waves are the analog of the field in the
background metric determined both by the background flow of the water and the
varying depth of the water (since the surface wave velocity depends on depth).
In order to measure the depth we needed to accurately measure the surface of
the water. We did this by dissolving the dye, rhodamin-C in the water and
illuminating a narrow strip of the water with a green .5W laser whose beam was
spread out to a length of about 2m along the surface and width of about 1mm.
The rhodamine-C had a sufficient density in the water than the mean path of
the light in the dyed water was only about 1mm. After absorbing the light, the
dye fluoresced with a broad peak below the frequency of the green laser. Since
the fluorescence was isotropic, this produced light in all directions,
including almost perpendicular to the laser light beam. (The laser light
itself tended to either specularly reflect at the surface or refract, neither
of which produced light in the perpendicular direction). The fluorescence also
destroyed the temporal and spatial coherence of the emitted light, and thus
 did not suffer from the
``speckle" problem that visualization under laser light usually produces. 

The bright surface fluorescent emission was then photographed with a digital
camera (BW to obtain the maximum resolution that the number of pixels could
produce, and with a 1980x1094 resolution) so that the full 2m illumination by
the laser could be recorded. 

Because of the softness of the lens (the  focal resolution of the lens would
smear out a point source of light to a size slightly larger than one pixel),
one could determine the location of maximum brightness in the image to much
better than 1 pixel by interpolation of the peak intensity from the pixels
immediately adjacent to that brightest pixel. (This gave a resolution of about
1/5 of a pixel. Since the pixels themselves had a size of about 1mm when
projected onto the water's surface, this gave a single pixel resolution from
the photographs of about .2mm in the vertical direction. Horizontal resolution
was not as important since the wavelengths of the waves of interest were of
the order of 10s of cm. After averaging by taking Fourier transforms of the
surface waves, we could reliably detect waves on the surface with amplitudes
down to about .01mm.

This surface resolution was important in that it allowed us to use very low
amplitude waves in our experiments to ensure that we remained within the
linear regime of wave propagation. As is well known, waves propagating up a
shoaling beach have their amplitude amplified because of the decreasing
velocity of the waves. (See for example the height and breaking  of tsunami waves as they
hit land, compared to their few cm height in the deep open ocean). 

The experiment we carried out was not on the analog of black hole horizons,
but rather on the analog of white hole horizons. White hole horizons are the
time reverse of black hole horizons. Whereas a black hole horizon is a surface
out of which no waves can come, a white hole horizon is one into which no
waves can penetrate. Since physics is time symmetric, the physics of, and the
quantum emission by, white hole horizons is the same as the time inverse of
black hole horizons. However, while in at a black hole horizon, the quantum
Hawking process creates low frequency, long wavelength outgoing modes, for a
white hole horizon, the particles created are the time inverse. As argued
above, the originating modes which create the Hawking radiation in the past
were ultra high frequency, ultra short wavelength modes. In the inverse
Hawking process it is these modes which are created. Fortunately in the analog
systems, the dispersion relations ensure that these modes  do not have the
absurdly high frequencies, or absurdly short wavelengths that they do for the
black holes. 

Thus the creation process for modes of the white hole analog horizons in our
experiment will create modes with short wavelengths, which turn out to be about
20cm in our case. The corresponding black hole modes would have wavelengths of
many 10s of meters which would be extremely difficult to measure in our 6m
tank, of which only 2m was illuminated. 

Thus, we had a wave generator, which consisted of a wire screen which was
immersed more or less deeply into the water flow downstream of the obstacle. This
screen would more or less impede the flow, producing a wave which would travel
upstream toward the obstacle. If the water flow depth was appropriately
adjusted ( by means of the vertical weir-- a adjustable vertical plate-- at the
end of the tank) then the upstream travelling waves would not travel over the
barrier. They were blocked. The obstacle would both slow down the waves (
whose wavelength was sufficiently long that their velocity was $\sqrt{gh}$
where here $g$ is the acceleration of gravity and $h$ is the depth of the
water). As the water shoals, the wave velocity decreased ($h$ get smaller)
while the velocity of the water increased (due to the incompressibility of the
water and the  conservation of mass for the water flow, the
velocity of the water times the depth is essentially constant). Thus, if the
depth of the water is properly
adjusted by the weir at the end of the tank, the velocity of the waves is
 less than the velocity of the water
over the obstacle, and no waves could penetrate the region  over the
obstacle. The waves are blocked. At the
``blocking point", the point where the group velocity of the waves equalled the
speed of the water, the waves cannot simply disappear. Instead, as in a
black hole, they pile up there, with their wavelength steadily  decreasing.
The dispersion relation of the surface waves 
\bea
\omega =\sqrt{ gk\tanh(kh)}
\eea
(where $k={2\pi\over \lambda}$ is the wave number and $\omega$ is the angular
temporal frequency of the surface waves) means that when the wavelength became small
enough, the group velocity of the waves will drop  below the velocity of the
water and the waves are swept away from the blocking point. 

If we assume that the fluid flow is steady (time independent) then, for the
small linear waves travelling over the surface of the fluid, the frequency of those waves
in the lab frame will be constant. With the dispersion relation in the still
fluid given by the above, the dispersion relation in flowing fluid will be
\bea
\omega = \sqrt{ gk\tanh(kh)} - vk
\eea
While this equation assumes a constant velocity $v$, it should also be a
reasonable approximation as long as $v$ does not change too fast. Thus at any
point in the flow, there will in general be three possible values of $k$ for
any small enough value of $\omega$ as long as v is not too large. In figure
4 we plot the above dispersion relation for two values of $v$
corresponding to different locations in the flow. In the slowly flowing fluid,
we have chosen a value of $\omega$ such that there are three possible values
of $k$. The smallest value  $k^+_i$ has a phase velocity $v_p={\omega\over k}$
and $v_G= {d\omega\over dk}$, the slope of the curve, which are both positive
( taken in this case to refer to velocities to the left). This corresponds to the
long wavelength ingoing wave. The other two solutions  in the slow water
regime $k_o^\pm$ with much shorter
wavelengths, both have negative slopes ( negative group velocities) which
correspond to waves dragged  away from the horizon. $k^+_o$ 
has positive phase velocity ($\omega$ and $k^+_o$ are both positive) while
$k^-_o$  has negative phase velocity. 

\begin{figure}\begin{center}
    \includegraphics[width=110mm]{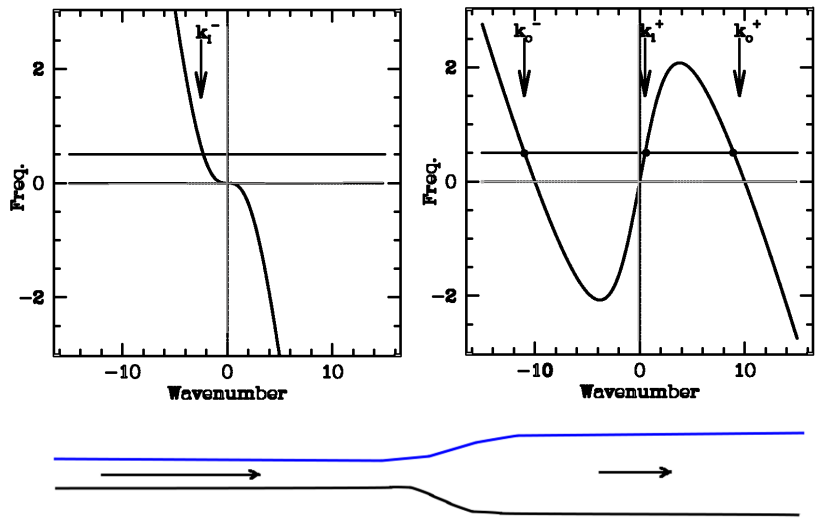}
    \caption{ The dispersion relations  for the surface waves in  regions 
where the flow is faster and slower  than the speed of the long wavelength surface waves. 
 For a given frequency, designated by the horizontal line above zero, $k^+_i$ is the positive norm long wavelength wave whose group
 velocity is  incoming toward the white hole horizon in
the slow flow region, while $k_i^-$ is the negative norm mode whose group velocity
is incoming toward the horizon in the fast flow region. 
$k_o^+$ and $k_o^-$ are waves whose group velocities carries them away from the horizon by the flow, and
represent the waves which are created by the horizon from the incoming waves
$k_i$. The wave $k_o^-$ is a negative norm wave, and $k_o^+$ a positive norm wave,
and their intensity ratio is the ratio of Bogoliubov coefficients, and should
have a thermal character if the Hawking analysis is correct. In our
experiment, the mode corresponding to $k_i^+$ is generated, and the amplitudes
of the  resultant
waves $k_o^\pm$ are measured.}
    \end{center}\end{figure}

If we look at waves in the shallow, high-velocity region, there is only
one solution to the dispersion relation with that same value of $\omega$. That
wave has negative group and phase velocities,-- i.e., directed to the right  toward the horizon.
Thus, 
both this wave $k_i^-$ ,  and the long wavelength possibility in the deeper
water regime
$k_i^+$ 
represent waves travelling toward the horizon, while both the short wavelength
waves in the deeper water are travelling away.

When one send a long wavelength wave at the horizon from the right, it will
eventually be stopped by the flow, and be converted into the two outgoing
short wavelength waves. No outgoing wave  can enter into the fast flow region, because
there is no solution there with group velocity away from the horizon. (there
do exist waves with imaginary wave-number  there of course, which would
correspond to  exponentially damped solutions and  which would in general be
needed to satisfy the boundary conditions at the horizon.)

What we measured in our experiment was precisely that conversion
of the ingoing waves into outgoing waves. In particular the amplitudes of those
various outgoing waves was the crucial output of this experiment. 

\section{Norm}

Before continuing with the description of the experiment, I must return to
quantum mechanics. The above experiment sounds completely classical. But the
Hawking effect is surely a quantum effect-- $\hbar$ occurs in the formula for
the temperature. How could this classical experiment have anything to do with
quantum mechanics?

For any linear system\cite{unruh-amplifier}  (i.e., a system with a quadratic Hamiltonian), as those
surface waves are, there is a conserved norm for complex solutions of the wave
equations. If $\phi_i$ and $\pi_i$ are the field variable and conjugate momentum,
then 
\bea
(\tilde q,q)={i\over 2}\sum_i\left( \tilde \pi_i q_i -\tilde q_i \pi_i\right)
\eea
defines  an inner product between the solutions $\tilde q_i,\tilde pi_i $ and
$q_i,\pi_i$. which is conserved in time even if the Hamiltonian is explicitly time
dependent. 

This inner product can be used to define a norm for complex solutions
\bea
<q_i,q_i>= {i\over 2}\sum_i\left( \pi_i^* q_i -q_i^* \pi_i\right)
\eea
since the Hamiltonian is real, and thus if $q_i,\pi_i$ is a solution, so is
its complex conjugate.

This norm is not positive definite, and the norm of real solutions is zero.
However one can choose a set of solutions $\{\hat q^x_i\}$ which have positive norm,
are orthogonal to each other and to the associated set of complex conjugate
solution, and such that the whole set of solutions are a complete set of
solutions. We will call this set the positive and negative norm mode
solutions. If the $q_i$ are normalized ($<q^x,q^y>=\delta^{xy}$, and we define
annihilation and creation operators $a^x,~a^{x\dagger}$ such that 
\bea
[a^x,a^{y\dagger}]=\delta ^{xy}
\eea
then the quantum operators
\bea
Q_i= \sum_x a^x q^x_i +a^{x\dagger} q^{x*}_i\\
\Pi_i= i\sum_x( -a^x q^x_i +a^{x\dagger} q^{x*}_i)
\eea
obey the standard equal time commutation relations
\bea
[Q_i(t),\Pi_j(t)]=i\delta_{ij}
\eea

Ie, these operators obey the Heisenberg equations of motion for the
Hamiltonian (because the solutions $q^x_i$ all do) and obey the commutation
relations for configuration and conjugate momentum. 

If we know what the classical solutions of the equations of motion are we
also know what the full solutions to the quantum system are. 

\section{Results}

The Bogoliubov coefficients are the relation between the ingoing modes and the
outgoing modes. In particular, if and ingoing positive norm mode $\phi_i$ is
converted into a linear combination of outgoing positive and negative norm
modes $\phi^+_o,~\phi^-_o$ such that 
\bea
\phi_i\rightarrow \alpha\phi^+_o+\beta\phi^-_o
\eea
where all three have unit (positive or negative) norm, then because of the
conservation of norm, $|\alpha|^2-|\beta|^2=1$ and the ratio 
\bea
{|\beta|^2\over| \alpha|^2}= e^{-{\hbar\omega\over K_B T}}
\eea
defines the temperature of the emitted quantum radiation. If
$ln({|\beta|^2\over| \alpha|^2})$ is linear in $\omega$ the temperature is a
constant, and in Hawking's calculation, related to the properties of the
horizon. 

\begin{figure}\begin{center}
    \includegraphics[width=80mm]{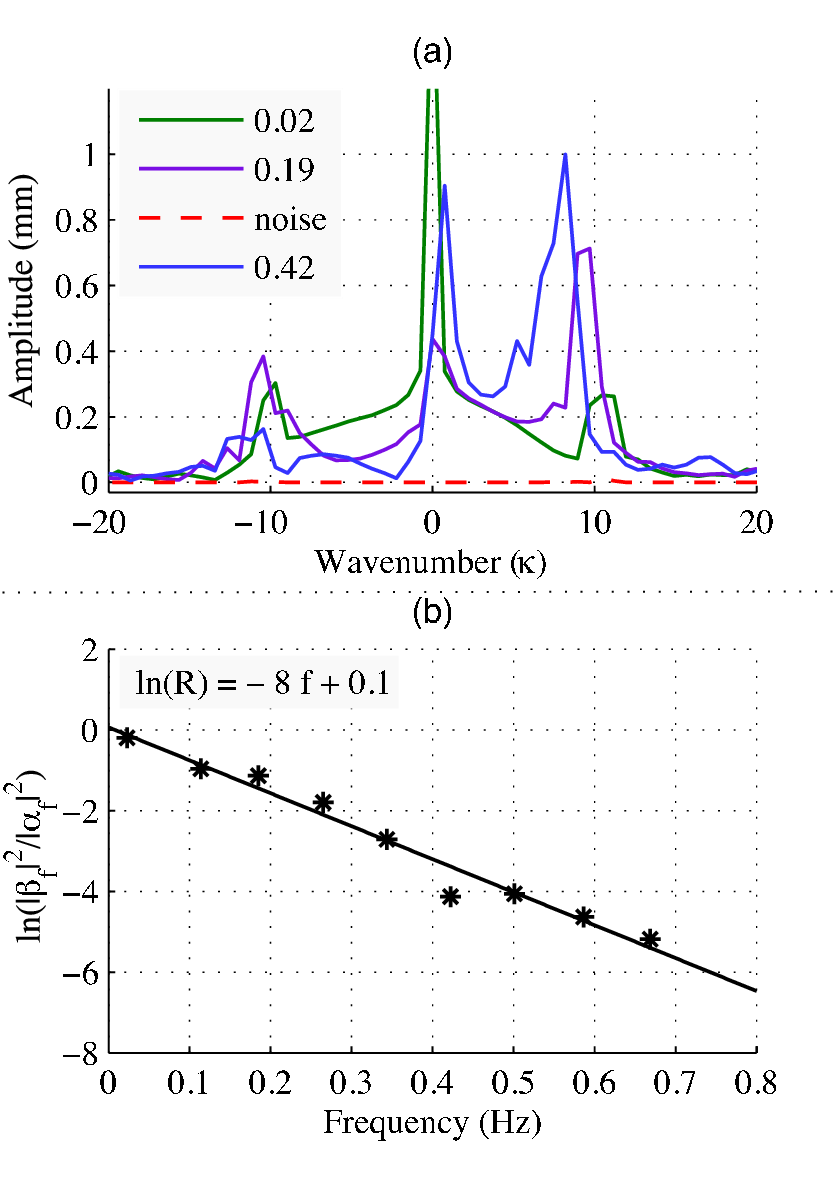}
    \caption{ The plot\cite{experiment}  of the intensities of the radiation as a
function of wave-number for the waves incident on the barrier. The peak near
$k=0$ is the incoming wave, while the two peaks, one at positive and one at
negative $k$ represent the positive and negative norm waves produced by the
interaction of the incoming wave with the horizon. Note that the wavelengths of
the incoming waves are much longer than the illuminated region of the top of
the water, the waves are travelling over an uneven bottom, and there is no
wavelength matching of the analyzed region, making the peaks broad, even
though only one frequency was incoming. The logarithm of the ratio of the
intensities is plotted in the other graph. If the thermal hypothesis is
correct, then that plot should be a straight line, which, to experimental
accuracy, it is. Unfortunately the temperature corresponding to the slope of
the graph $T={h\nu_0\over k_B }$ where $\nu_0$ is the inverse slope,
corresponds to a temperature of about $10^{-12}K$, slightly cooler than the
water we used in the experiment. }
    \end{center}\end{figure}

Thus, this experiment give strong support to the hypothesis that horizons,
whether black hole, sonic, or other will produce a quantum noise with a
thermal spectrum, whose temperature is determined by the behaviour of the
horizon. 

\section{Future}

What remains? Clearly it would also be good to be able to directly see the
thermal quantum noise created by a horizon. As in the above experiment, this
is extremely difficult. The temperature scales directly with the velocity of
the waves, and inversely with the scale over which the horizon is created.
This suggests that horizons which are created by light (high velocity and thus
high temperature)  are preferable, and also 
horizons which are created over very short length scales are prefered. One  suggestion
is that one look at non-linear effects in optical media to create horizons.
Schuetzhold and I\cite{schuetz-waveguide}  made  described one possibility in
which the effective velocity of electromagnetic waves in a wave guide could be
altered and used to create a (moving) horizon. This  was
extended  by Leonhardt\cite{leonhardt}  who
suggested using the non-linear Kerr effect ( the change in the refractive
index of a transparent medium by an intense pulse of radiation in the medium).
One would use an intense pulse at one frequency  to change the refractive
index for a different frequency such that at that other frequency, the
velocity of the pulse was higher than the velocity of the light within that
pulse. This would create a black hole/white hole pair of horizons in the rest
frame of the pulse. This is  probably the experiment  which is closest to fruition,
but still faces immense obstacles. The because of the weak non-linearities of
any known medium, the intensity of the pulse which changes the refractive
index  need to be so high 
that it begins to damage the material if it is to create a sufficiently large
change in the index of refraction within the pulse. This damage can created
 radiation noise in a broad range of
frequencies\cite{faccio}. However, the hope is that
in the next 10 years or less, the first direct detection of radiation created
by an analog horizon will have been seen\cite{analogbook}.  

Furthermore in the analog experiment which we carried out, the behaviour of the
waves was in some sense too predictable. The modes  were all in a regime in
which the physics is reasonably well understood (if we ignore turbulence and
viscose effects). It would be great if one
could carry out an experiment where at least some of the modes  were in a regime in which the
physics was poorly understood. For example, in liquid He, sound waves, or
rather modes of vibration,  whose
wavelengths are comparable to the inter-atomic spacing in the liquid are very
poorly understood.    If one could run an experiment in which the black hole
horizon radiation originated with modes which lay within that region of
atomic wave-length ``sound waves", it would strengthen the evidence that the thermal radiation
from horizons really was as ubiquitous as it seems to be. Unfortunately such
experiments still seem a long way off. 

It is of course true that even the direct observation of thermal quantum
radiation from an analog horizon does not prove that black holes will radiate.
Something could make the gravitational system behave differently from any
analog system. It is however very hard to imagine what that something could
be. The derivation of the thermal radiation from analog horizons follows so
closely the derivation of thermal radiation  from  black hole horizons that it
is very hard to imagine how the one could occur but not the other.

However our  first experimental demonstration that  a horizon produces a
thermal spectrum
 together with  the very elementary arguments that, for linear systems, the
classical behaviour determines the quantum behaviour,  
is at least a first step to solidifying the truth  of
Hawking's observation that horizons are associated with thermal radiation,
despite the problematic nature of his original derivation.  

\section{Ackowledgements}
I would like to thank NSER of Canada, CIfAR, and the Templeton Foundation for
support of this research. I would also like to thank the  Perimenter Institute
and through them also the taxpayers of Ontario and Canada through their
governments who fund it for
 hosting and supporting  me during part of the time this research was carried out.

\end{document}